# Nowcasting ETAS Earthquakes:
# Information Entropy of Earthquake Catalogs

by

Ian Baughman[1]
John B Rundle[1,2,3]
Tianjian Zhang[1]

[1] Department of Physics and Astronomy
University of California, Davis, CA USA

[2] Department of Earth and Planetary Sciences
University of California, Davis, CA USA

[3] Santa Fe Institute
Santa Fe, NM, USA

## Abstract

Earthquake nowcasting has been proposed as a means of tracking the change in large earthquake potential in a seismically active area.  The method was developed using observable seismic data, in which probabilities of future large earthquakes can be computed using Receiver Operating Characteristic (ROC) methods.  Furthermore, analysis of the Shannon information content of the earthquake catalogs has been used to show that there is information contained in the catalogs, and that it can vary in time.  So an important question remains, where does the information originate?  In this paper, we examine this question using statistical simulations of earthquake catalogs computed using Epidemic Type Aftershock Sequence (ETAS) simulations. ETAS earthquake simulations are currently in widespread use for a variety of tasks, in modeling, analysis and forecasting.  After examining several of the standard ETAS models, we propose a version of the ETAS model that conforms to the standard ETAS statistical relations of magnitude-frequency scaling, aftershock scaling, Bath's law, and the productivity relation, but with an additional property.  We modify the model to introduce variable non-Poisson aftershock clustering, inasmuch as we test the hypothesis that the information in the catalogs originates from aftershock clustering.  We find that significant information in the catalogs arises from the non-Poisson aftershock clustering, implying that the common practice of de-clustering catalogs may remove information that would otherwise be useful in forecasting and nowcasting.  We also show that the nowcasting method provides similar results with the the ETAS models as it does with observed seismicity.

## Plain Language Summary

Earthquake nowcasting has been proposed as a means of tracking the change in the potential for large earthquakes in a seismically active area, using the record of small earthquakes. The method was developed using observed seismic data, in which probabilities of future large earthquakes can be computed using machine learning methods that originally were developed with the advent of radar in the 1940s. These methods are now being used in the development of machine learning and artificial intelligence models in a variety of applications. In recent times, a method to simulate earthquakes using the observed statistical laws of earthquake seismicity has been developed, the Epidemic Type Seismicity Sequence "ETAS" model. One of the advantages of the ETAS model is that it can be used to analyze the various assumptions that are inherent in the analysis of seismic catalogs of earthquakes. In this paper, we analyze the importance of the space-time clustering that is often observed in earthquake seismicity. We find that the clustering is the origin of information that makes the earthquake nowcasting methods possible. We find that a common practice of "aftershock de-clustering", often used in the analysis of these catalogs, removes information about future large earthquakes.

## Key Points

- Earthquake nowcasting was proposed as a means of tracking the change in the potential for large earthquakes in a seismically active area
- We analyze the information contained in the space-time clustering that is often observed in earthquake seismicity
- We find that "aftershock de-clustering of these catalogs removes information about future large earthquakes.

## Introduction

Earthquake nowcasting (Rundle et al., 2019; 2021a,b,c; 2022; 2023; Rundle and Donnellan, 2020; Pasari 2019; 2020; 2022; Pasari and Mehta 2018; Chouliaris et al., 2023) is a relatively new method that uses elements of machine learning to track the current state of the potential for large earthquakes, as well as the recent past and near future. It is based on the Receiver Operating Characteristic (ROC) method that was developed with the invention of radar in 1941 (https://en.wikipedia.org/wiki/Receiver_operating_characteristic) relating to the observation of "signals" associated with reflections, or "events".

In the nowcasting application, we compute the current large earthquake potential state (EPS), which is defined as the exponential moving average (EMA) of the inverted monthly rate of small earthquakes. We then define a "signal", which is the current value of the EPS, and an "event", which is the occurrence or non-occurrence of a large earthquake within a selected future time window $T_W$. In (Rundle et al., 2022; 2023), the value of $T_W$ is typically 1-3 years.

To implement the ROC method, we establish an arbitrary threshold value for the EPS curve, and categorize the "signals" and "events" into 4 categories, thereby building a "confusion matrix" or "contingency table". A "signal" is a value of the EPS curve either above or below the value of the arbitrary threshold.

There are four possibilities. 1) If the "signal "is above the threshold, and a large earthquake *does occur* within the following $T_W$ years ("event"), the "signal" is a True Positive (TP). 2) If the "signal "is above the threshold, and a large earthquake *does not occur* within the following $T_W$ years ("event"), the "signal" is a False Positive (FP). 3) If the "signal "is below the threshold, and a large earthquake *does occur* within the following $T_W$ years ("event"), the "signal" is a False Negative (FN). 4) If the "signal "is below the threshold, and a large earthquake *does not occur* within the following $T_W$ years ("event"), the "signal" is a True Negative (TN).

The initial threshold value is established at the lowest value of the EPS curve, all points on the timeseries curve are evaluated, and the confusion matrix is built for that value of the threshold. The threshold is then incrementally increased, and a new confusion matrix is built for that new threshold. This process continues until the maximum value of the EPS curve is reached. Typically one chooses several hundred threshold values, each resulting in its characteristic confusion matrix. From these confusion matrices, the ROC curve is then constructed by plotting the True Positive Rate (TPR), against the False Positive Rate (FPR), where:

$$TPR = \frac{TP}{TP + FN}, \quad FPR = \frac{FP}{FP + TN} \qquad (1)$$

From these quantities, we can also compute the Positive Predictive Value, or Precision (PPV), which is defined as:

$$PPV = \frac{TP}{TP + FP} \qquad (2)$$

The precision can be regarded as the probability of a future large earthquake occurring within the following $T_W$ years if the EPS curve is at, or above, a reference value on the EPS curve.

As discussed by Rundle et al. (2023), the fact that one can compute the PPV value implies predictability for large earthquakes. This in turn implies that there is Shannon information (Shannon, 1948) contained within seismic catalogs. An important question is, from where does this information arise? What property of the catalogs is associated with this information content? This is the question that we seek to answer in the current paper.

The hypothesis that we consider here is that the information arises from the non-Poisson clustering of the earthquake data, typically as a result of aftershocks following large earthquakes, or as swarms or "bursts". More specifically, we find that information is associated with the relative quiescence that follows aftershock activity, meaning that removing aftershocks obscures the boundary between an "active" phase vs. a "quiescent" phase.

To test this hypothesis, we turn to earthquake simulations in the form of Epidemic Type Aftershock Sequences (ETAS) models. These models are constructed using the standard statistical relations of magnitude-frequency scaling, aftershock scaling, Bath's law, and the ETAS productivity relation. We then apply the nowcast method to these ETAS models and discuss the results.

## Epidemic Type Aftershock Sequence (ETAS) Models

**Overview of Current ETAS Models.** The Epidemic-Type Aftershock Sequence (ETAS) is a statistical model for the spatial and temporal distribution of seismic activity for a given geographical region. Apparent from the name, the model's focus is on aftershocks and capturing the observed clustering following a mainshock. The framework that the ETAS model provides has found use beyond modeling single aftershock sequences, to compute catalogs of events (Zhuang et al., 2015). Once the model has been calibrated for a specific spatial and temporal region it can be used in forecasting seismic activity in various seismically active regions (Zhuang, 2011; Mancini et al., 2021).

The ETAS model focuses on the clustering of aftershocks following mainshocks. It does this by differentiating between two different types of events: background and triggered. The production of events is treated as a branching process where background events occur spontaneously due to some external loading process (i.e. tectonic loading, subsurface fluid injection, etc.) that have the potential to trigger future events. It is important to note that triggered events can themselves trigger future events leading to self-exciting cascades.

Mathematically, this is expressed as a conditional intensity.

$$\lambda(M, x, y, t) = s(M)\left[\lambda_0(x, y) + \sum_{\{i:\, t_i < t\}} \kappa(M_i) g(t - t_i) f(x - x_i, y - y_i; M_i)\right] \quad (3)$$

Technically speaking, this is a "marked" branching process, meaning that every event that occurs is marked with a magnitude, $M$. This magnitude is determined by $s(M) = \beta e^{\beta(M-\mu)}$, the probability density function (pdf) form of the Gutenberg-Richter distribution. Here $\mu$ is the completeness magnitude and $\beta = b \ln 10$ is related to the $b$-value. When and where events occur is determined by the expression in the brackets.

Background events in this model are assumed to follow a Poisson distribution (Gardner and Knopoff, 1974) with $\mu(x, y)$ being the rate (i.e. the background rate) which is constant in time but allowed to vary spatially. The summation term describes the triggering process. Here $\kappa(M) = K e^{\alpha(M-\mu)}$ is the productivity relation, it represents the mean number of offspring of an event with magnitude $M$.

The terms $g(t - t_i)$ and $f(x - x_i, y - y_i; M_i)$ are density functions for the occurrence time and location of an event triggered by the $i$-th event, respectively. Various functional forms have been used for $f(x - x_i, y - y_i; M_i)$ (Zhuang at al., 2011) while $g(t - t_i)$ is taken to be the probability density form of the Omori-Utsu Law:

$$g(t) = \frac{p-1}{c}\left(1 + \frac{t}{c}\right)^{-p} \quad (4)$$

Before implementing this model, one must first determine "reasonable" model parameters. We say "reasonable" because certain combinations of model parameters can lead to a critical state where cascades of triggered events will fail to terminate. For more information on criticality in the ETAS model refer to (Zhuang et al., 2012). Often though, the ETAS model is employed in the study of seismicity for a specific geographical region and time window where an empirical catalog already exists.

To determine model parameters which best describe the desired catalog, a log-likelihood function is chosen and maximized to give a maximum likelihood estimator (MLE) of the model parameters, $\hat{\theta}$. For a review of the log-likelihood function refer to (Zhuang et al., 2011; 2012, 2015) and (Veen et al., 2008; Lombardi, 2015) for computational methods to maximize the log-likelihood function Another example is the recent paper by Mancini and Marzocchi (2023) that discusses how to fit an ETAS model ("SimplETAS") to a given region. Still another is the paper by Hardebeck (2013).

While the ETAS model performs well at modelling earthquake catalogs, the determination of model parameters is far from a trivial task. The spatial and temporal windows chosen, as well as the completeness magnitude used, can all lead to substantial variations in the MLE parameters (Seif et al., 2017). Furthermore, the notion of an aftershock within the ETAS model has some ambiguity, in the sense that it is not so clear when the aftershock cascade terminates and the events become "background".

Whether we choose an aftershock to be direct offspring of an event or the entire sequence of triggered events (i.e. directly triggered offspring plus indirectly triggered) can lead to deviations in the observed Omori-Utsu $p$-value. If instead, the aftershocks are determined from the ETAS catalog using conventional methods usually employed for empirical catalogs, still different results are obtained (Helmstetter and Sornette, 2003).

**Modified ETAS Model with Variable Clustering.** In a modified model that we consider here, we start with the same basic equations as the current ETAS models. Specifically, our basic ETAS equation for the rate of earthquake production is an implementation of the standard approach:

$$\lambda(M,x,y,t) = s(M)\left[\lambda_0(x,y) + \sum_{t_i<t} \frac{K 10^{\alpha M_i}}{(\Delta t_i + \tau)^p} \frac{1}{(\Delta r_i + \xi)^{2q}}\right] \qquad (5)$$

where:

$$\Delta t_i = t - t_i \qquad (6)$$

$$\Delta r_i = \sqrt{(x-x_i)^2 + (y-y_i)^2)} \qquad (7)$$

Parameters $\lambda_0$, $K$, $\alpha$, $p$, $q$, $\tau$, and $\xi$ are the usual parameters in ETAS models, $t$ is time, and $(x_i, y_i)$ are the spatial locations of the ETAS earthquakes (Table 1). Furthermore, $\mu$ is the magnitude of the smallest earthquake (catalog completeness magnitude), and $M_i$ is the magnitude of the event at time $t_i$. All parameter values used in this paper are given in Table 1. Of course magnitudes are picked from a Gutenberg-Richter distribution with parameter $b$.

To build catalogs, we start with a relation to compute the "bare" or "unrenormalized" time to the next earthquake. Later we describe how the variable clustering (renormalized clustering) is computed. The (Poisson) equation to determine the time to the next earthquake is the following, where $T_S$ is a time scale:

$$\Delta t = - T_S \, R \, (Log(1 - R * \zeta)) \tag{8}$$

and where:

$$R = \left(\left(\frac{\lambda(\mu,x,y,t)}{\lambda(M,x,y,t)}\right)\right)^{\eta} \leq 1 \tag{9}$$

Here: $\zeta$ = Random number drawn from a uniform distribution. Note in particular that the equation for $\Delta t$ arises from the inversion of a Poisson cumulative distribution function, and that the natural $Log$ term in parenthesis is $\leq 0$. The role of $R$ can be seen to reduce the time interval to the next earthquake after a large earthquake occurs.

In our implementation here, the number of events in the simulation catalogs is chosen to be the same as for the California catalog, given the same completeness magnitude $\mu$. As we describe below, the time scale $T_S$ plays no important role since all event times will be proportionately rescaled so that the resulting catalog is the same total duration as the observed California catalog, thus matching the long term rate of seismicity in the region.

For the spatial locations of events, we use the following algorithm. We first determine whether an event is a random "background" event, or whether it arises as an "aftershock" of a previous "sufficiently large" event. To determine this difference, we establish a counter, $\Phi$, which is set to zero, $\Phi = 0$, at the beginning of the catalog simulation.

If no "sufficiently large" earthquake occurs as the next event, the counter is decremented by 1, i.e., $\Phi \to \Phi$ -1. At negative values of the counter $\Phi$, all earthquakes are considered to be "background" events. At positive values of counter $\Phi$, events are considered to be "aftershocks".

By "sufficiently large", we mean an earthquake whose magnitude is:

$$M > \mu + \Delta M_B \tag{10}$$

where $\Delta M_B$ is the Båth's law constant $\sim 1.2$, the average difference between a mainshock and its largest aftershock.

If such a large enough earthquake does occur, we compute the expected number N of aftershocks:

$$N = 10^{M - \mu - \Delta M_B} \tag{11}$$

We then set the counter $\Phi = N$, and continue this process, computing the time to the next aftershock, until the counter decreases to a value below 0, unless a sufficiently large aftershock having magnitude occurs:

$$M_A > \mu + \Delta M_B \qquad (12)$$

If such a large aftershock does occur, we compute the expected number $N_A$ of events from that aftershock:

$$N_A = 10^{b(M_A - \mu - \Delta M_B)} \qquad (13)$$

If at that point, $N_A > \Phi$, we reset $\Phi = N_A$, otherwise we continue with the decrementing process. By this mechanism, we allow large aftershocks to extend the cascade of aftershocks beyond what would occur only from the mainshock. Thus, aftershocks can have aftershocks as well. Once the aftershock process is complete, and there is no further extension of the aftershock count, the counter $\Phi$ reaches zero and continues into negative numbers if no new aftershock occurs that is large enough to also produce aftershocks.

To choose the locations of the background events, we pick the epicenters at random from the existing catalog of locations of observed earthquakes. For large mainshocks, we pick from the epicenters of the few largest earthquakes in the catalog. For smaller mainshocks, we pick from any epicenter in the catalog.

For the aftershock locations, once the counter is positive, we carry out a random walk with step size $\delta$ measured in degrees, around the epicenter of the mainshock, until the counter reaches zero. We then revert to picking locations at random from the existing catalog of observed locations. In this paper, we have no need for the spatial locations, so this aspect will not be discussed further.

**Enhanced (Non-Poisson) Clustering of the Catalog.** Once the basic catalog is built, we scale the times to the same interval as the USGS catalog, from 1960 to the present. This is done by building an ETAS catalog with the same number of events, with the same completeness level, as the USGS catalog for California. We then scale all the times of the ETAS events in the catalog to the time interval 1960-present.

We enhance the clustering of the aftershocks so as to reproduce the observed temporal scaling with an Omori-Utsu p-value near 1, as observed in nature. This process involves the use of a geometric scaling factor $f < 1$, so that scale-invariant clustering is produced. The factor $f \equiv 1 - \varepsilon$, where $\varepsilon$ is a small (positive) random number drawn from a uniform distribution between 0 and a maximum value of $\varepsilon$.

We choose a clustering factor $f < 1$ and apply it recursively across all magnitude scales. We identify "cycles" of mainshocks and their inter-event "aftershocks". We then apply the clustering algorithm to the "aftershocks". We begin with the smallest earthquake that can have aftershocks and apply the clustering algorithm recursively to all cycles of increasing magnitude up to the largest possible aftershock, which is determined using Båth's law. This moves all the "aftershocks" geometrically closer to the first "mainshock", with the exception of the last "aftershock".

As mentioned above, some of the current ETAS models can demonstrate aftershock cascades that fail to terminate. In addition, there can be differences in Omori-Utsu p-values that are seen in the directly triggered aftershocks, and the indirect aftershocks. Both of these are problems that do not occur in our modified ETAS model, an important constraint since our hypothesis is that it is the aftershocks, or lack of them, that is the origin of the (Shannon) information entropy.

In general, the statistics of of the modified ETAS model are in basic agreement with the statistics of the observed data, although there can be inevitable small statistical variations between different computational runs using the same input parameters. In the figures below we show a comparison between the statistics of the USGS California catalog and our modified ETAS model.

## Nowcasting with California Earthquakes

In previous papers, we have described the nowcasting model, which tracks the current state of the complex dynamical earthquake system in space and time (e.g., ). Construction of a nowcast diagram uses the California earthquake catalog since 1960-present within $5^o$ latitude and longitude of Los Angeles, California. Details are discussed in Rundle et al. (2022), and the reader is referred to that paper for details.

Briefly, the method begins with the seismicity data in the seismically active region around Los Angeles, and applies an exponential moving average (EMA) to the monthly rate of small earthquakes in the region, using a number of weights N = 36 months, in this case. We also apply a correction to optimize the ROC skill of the nowcast, and to approximately correct for small earthquake counts not present in the catalog. In that method, we used earthquakes having $M \geq 3.3$ since 1960.

After applying the EMA and small earthquake correction, the result is a smoothed version of the monthly rate of small earthquakes, which resembles an "upside-down" version of the earthquake cycle of stress accumulation and release, as has been pointed out in Rundle et al. (2022). Inverting this smoothed seismicity curve, we obtain the nowcast timeseries curve, examples of which will be shown in the figures, in which seismic activation is at the bottom of the diagram (many earthquakes such as aftershocks), and quiescence is at the top. Thus we see that quiescence is the precursor to large earthquakes.

Figure 1 shows a number of important details, where we note in passing that the Gutenberg-Richter *b*-value was found to be *b* = 0.99. Figure 1a is a plot of the magnitude of earthquakes as a function of time, and clearly shows the clustering in time that small earthquakes are known to exhibit. This clustering is a combination of both expected "Poisson" clustering (Gross and Rundle, 1988) as well as "non-Poisson" clustering. Figure 1b is a plot of the stacked aftershocks from all the 104 earthquakes *M* > 5.5, showing a *p*-value of 1.02. To create this plot, aftershocks were identified and labeled using the procedure discussed previously for identifying ETAS aftershocks.

Figure 1c shows the nowcast, along with the middle plot (Figure 1d) showing the precision (Positive Predictive Value ≡ *PPV*), and the last plot at right (Figure 1e) showing the Shannon information entropy (Rundle et al., 2023). Both the precision and the information of the data are represented by the solid line. The 50 cyan curves in each of the Figures 1d and 1e are computed by randomizing the nowcast curve using a bootstrap procedure (random sampling with replacement), then computing the precision and information. Mean and standard deviation of the cyan curves are indicated by the dashed and dotted lines.

Also shown in Figures 1d and 1e are the maximum probability gain, MaxPG = 272.3%, and the maximum information gain, MaxIG = 1.897 bits. We define probability gain as:

$$\text{MaxPG} \equiv max(\ (PPV(\Theta)/<PPV_R(\Theta)>) - 1) \qquad (14)$$

where $PPV(\Theta)$ is precision, and $<PPV_R(\Theta)>$ is the mean precision of the random $PPV$ curves.

Self information $I$ is defined as (Shannon, 1948; Cover and Thomas, 1991):

$$I(\Theta) = -Log_2(PPV) \qquad (15)$$

where the information units are in "bits". Information gain is defined as:

$$\text{MaxIG} \equiv max(<I_R(\Theta)> - I(\Theta)) \qquad (16)$$

where $I(\Theta)$ is information entropy, and $<I_R(\Theta)>$ is mean information entropy. Both $PPV$ and $I$ are functions of the Earthquake Potential State ($EPS \equiv \Theta$).

In summary, it is clear from Figure 1c that the "precursor" to large earthquakes is anomalous or relative quiescence, which is the reason earthquakes are so hard to anticipate. Just as clearly, if one "de-clusters" the catalogs, as is common practice in many papers analyzing seismicity (e.g., Rundle et al., 2021a), the chance of detecting this anomalous quiescence will be reduced. In the following section, we provide an analysis of this problem.

**Nowcasting with ETAS Earthquakes**

To analyze the effect of non-Poisson clustering on the nowcast method, we begin by adopting a simplified version of our ETAS model as defined above. Specifically, we set the values $p = 0$ and $\alpha = 0$ in equation (5), which basically allows only Poisson clustering arising from equation (5). What remains is the non-Poisson clustering that arises as a result of the clustering process described previously.

We also assume that the constant-in-time background rate of activity $\lambda_0(x,y)$ = constant for simplicity. Other parameter values are described in Table 1. We consider two cases, the first (and baseline) case having $\varepsilon = 0.005$, the second case having $\varepsilon = 0.0005$. It will be seen that larger values of $\varepsilon$ are associated with increased non-Poisson clustering.

It should be emphasized that in both cases, for $\varepsilon = 0.005$ and for $\varepsilon = 0.0005$, the same non-clustered catalog was used as the base catalog, after which the clustering algorithm was applied to compute the resulting two catalogs. Following the computation of the clustered catalogs, the nowcasting process was then applied. Figures 2 and 3 shows the application of the same nowcasting method to clustered ETAS catalogs having $\varepsilon = 0.005$ and $\varepsilon = 0.0005$. The steps used in the applications are the same as were used in Figure 1 for the California catalog, as can be seen by comparing Figures 1-3.

Figures 2 and 3 show the magnitude-time plots (Figures 2a and 3a), and the Omori-Utsu aftershock decay from 128 aftershocks having $M > 5.5$ in Figures 2b and 3b. Figures 2c and 3c show the nowcast, Figures 2d and 3d show the precision $PPV(\Theta)$ as a function of earthquake potential state Θ, together with the mean precision from 50 randomized nowcast curves $PPV_R(\Theta)$. Figures 2e and 3e show the Shannon information $I(\Theta)$, as well as the information from the 50 randomized nowcast curves. Figures 2 and 3 should be compared directly to Figure 1.

Listed in these figures, we can see that MaxPG for California earthquakes is 236.4%, compared to values of 83.7% for $\varepsilon$ = 0.005, and 35.3% for $\varepsilon$ = 0.0005. The corresponding values of MaxIG is 1.75 bits for California; 0.877 bits for $\varepsilon$ = 0.005; and 0.436 bits for $\varepsilon$ = 0.0005. Although the MaxPG and MaxIG are not as great for the ETAS case with 0.005 as for California, this is primarily due to the fact that the random PPV is larger for the ETAS case than for California. But the trend of higher MaxPG for larger values of $\varepsilon$ is clear.

## Discussion

We emphasize that the point of our modified ETAS model is not to present a method to optimally fit California data, but rather to examine the effect of non-Poisson clustering in the catalog. In all models, Poisson clustering continues to exist. We also note that many other authors have discussed the importance of seismic quiescence, which clearly can only be identified in comparison to activation. These authors include Kanamori (1981); Weimer and Wyss (1994); Wyss and Haberman (1988); Chouliaris (2009); Katsumata (2011); Huang et al. (2021); Ben-Zion and Zaliapin (2020); Varotsos et al. (2011; 2014; 2020); Zaliapin and Ben-Zion (2022).

We have presented a simplified version of an ETAS model that exhibits a combination of both Poisson and non-Poisson clustering of earthquakes in Figures 2 and 3. In the first case, with $\varepsilon$ = 0.005, shown in Figure 2, the nowcasting results are similar to the results shown for the California catalog (Figure 1). Temporal clustering of events and Omori aftershock decay are similar as well. This is an example of machine learning applied to the earthquake problem, using both ETAS catalogs as well as the actual California catalog.

Comparing the results from the two ETAS catalogs (Figures 2 and 3), it can be seen that both precision and information content increase as non-Poisson clustering increases. This implies that de-clustering earthquake catalogs, as is often done today, removes information. The information evidently lies in the anomalous quiescence that follows the clustered events, so that removing the clustered events obscures the quiescence "precursor".

In conclusion, our results suggest that earthquake catalogs should not be de-clustered prior to analysis, so that information is not removed. We note that we plan to use ETAS models together with AI-enhanced models to explore the utility of AI and deep learning in the problem of anticipating earthquake hazard in future publications.

**Acknowledgements**. Research by JBR and IB was supported in part under DoE grant DE- SC0017324 to the University of California, Davis. JBR would also like to aknowledge generous support from a gift to UC Davis by John Labrecque.

**Open Research.** Python code that can be used to reproduce the simulation results of this paper can be found at a GitHub repository (DOI: https://zenodo.org/record/8423413. Data for this paper was downloaded from the USGS earthquake catalog for California, and are available there. Python code at (DOI: https://zenodo.org/record/7186635) can be used to download and model these data for analysis using the methods of Rundle et al. (2022).

**Data.** Data for this paper was downloaded from the USGS earthquake catalog for California, and are freely available there. The Python code mentioned above can be used to download these data for analysis.

**Table 1.** ETAS Parameters and Values

| ETAS Parameter | Description | Input Value |
|---|---|---|
| M | Magnitude | Variable |
| $\mu$ | Completeness Magnitude | 3.3 |
| $\Delta M_B$ | Båth's Law | 1.2 |
| K | Productivity Law | 1.0 |
| $\alpha$ | Productivity Law | Fixed at $\alpha = 0$ |
| q | Spatial Omori Law | 1.5 |
| p (input) | Temporal Omori Law | Fixed at p = 0 |
| p (observed) | Omori Aftershock Scaling | Variable Near 1.0 |
| b (input) | Gutenberg-Richter | 0.95 |
| b (observed) | Gutenberg-Richter | Variable Near 0.95 |
| $\varepsilon$ | Controls Recursive Scaling | 0.005, 0.0005 |
| $T_S$ | Time Scale | 0.1 days |
| $\tau$ | Correlation Time in Temporal Omori Law | 1 year |
| $\xi$ | Correlation Length in Spatial Omori Law | 100 km |
| $\lambda_0$ | Background Rate | Fixed = 5 |
| $\eta$ | Exponent in Rate Ratio | 1.5 |

**Figure 1.** Data for 15,838 California earthquakes with magnitude $M \geq 3.3$. a) Magnitude vs. time for California earthquakes from 1960 - present (9/2023). b) Omori-Utsu decay for 104 mainshocks having magnitudes $M \geq 5.5$. c) a) Shows the optimized state variable as a function of time, an enlarged version of Figure 1d. d) Shows the Positive Predictive Value, PPV or Precision. Magenta curve is the PPV for the state variable shown in c), where the vertical axis is the threshold $T_H$. The cyan lines represent the PPV for 50 random time series. Mean of the time series is the solid black line, and $1\sigma$ confidence is shown as the dashed lines. e). Magenta curve is the corresponding self information $I$, equation (15), on the horizontal axis as a function of the threshold value $T_H$ on the vertical axis. Again, the cyan curves are the self-information for the ensemble of 50 random time series, with mean (solid black line) and $1\sigma$ confidence as the dashed lines.

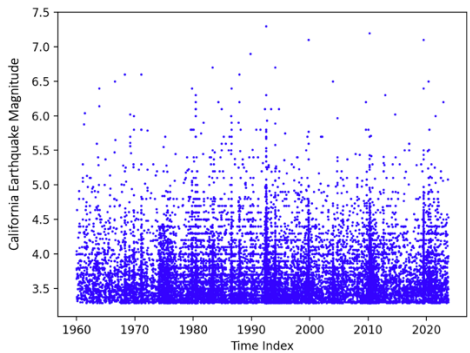

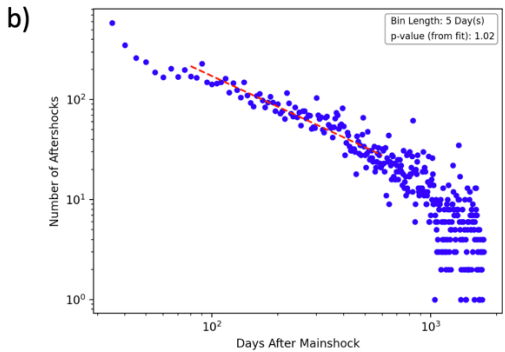

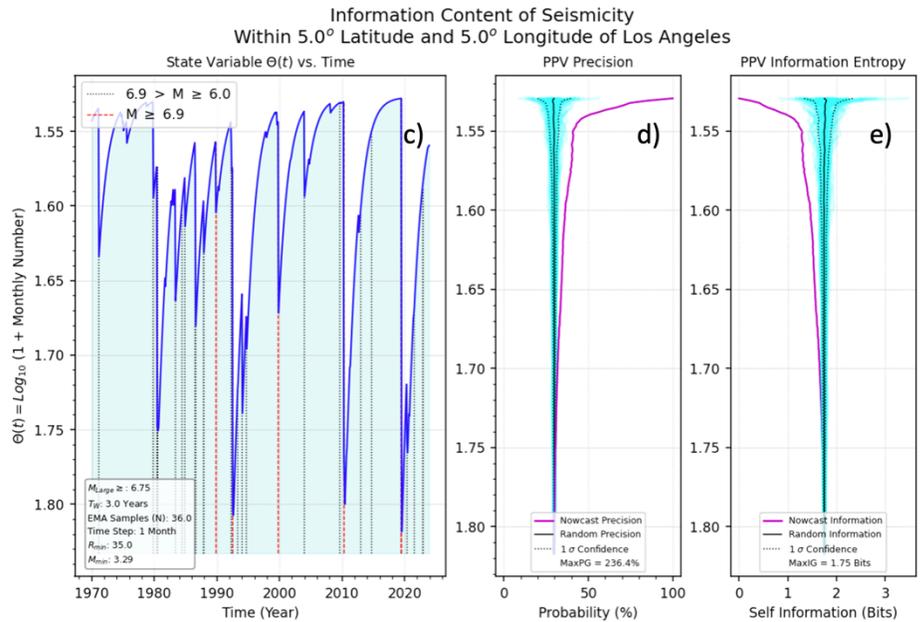

Observed Clustering with b = 0.98

**Figure 2.** Data for 16,001 ETAS California earthquakes with magnitude $M \geq 3.3$, and with a clustering factor $\varepsilon = 0.005$. a) Magnitude vs. time for ETAS California earthquakes from 1960 - present (9/2023). b) Omori-Utsu decay for 128 ETAS mainshocks having magnitudes $M \geq 5.5$, p-value over the interval shown by the red dashed line is $p = 1.12$. c) Shows the optimized state variable as a function of time, an enlarged version of Figure 1d. d) Shows the Positive Predictive Value, PPV or Precision. Magenta curve is the PPV for the state variable shown in c), where the vertical axis is the threshold $T_H$. The cyan lines represent the PPV for 50 random time series. Mean of the time series is the solid black line, and $1\sigma$ confidence is shown as the dashed lines. e). Magenta curve is the corresponding self information $I$, equation (15), on the horizontal axis as a function of the threshold value $T_H$ on the vertical axis. Again, the cyan curves are the self-information for the ensemble of 50 random time series, with mean (solid black line) and $1\sigma$ confidence as the dashed lines.

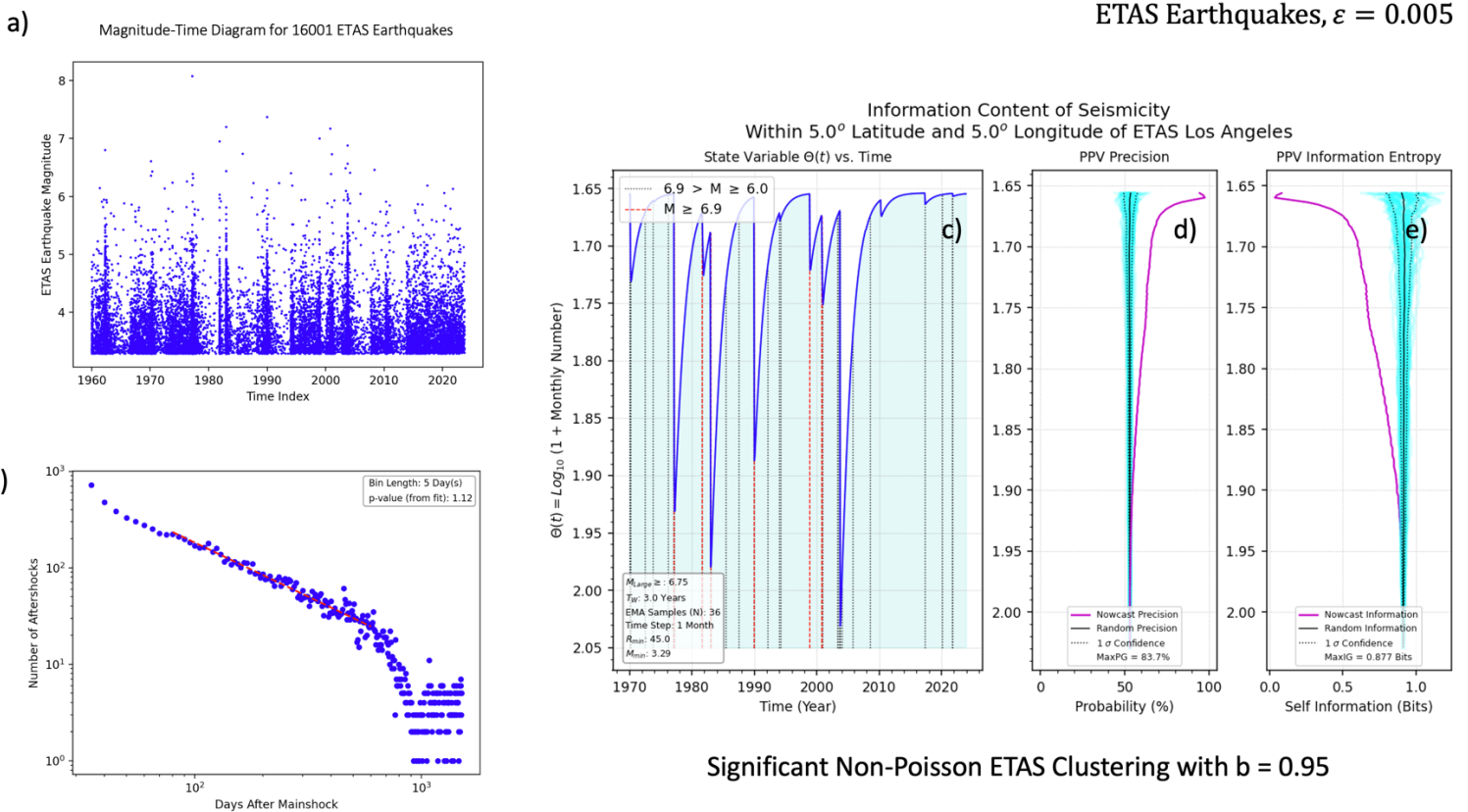

**Figure 3.** Data for 16,001 ETAS California earthquakes with magnitude $M \geq 3.3$, and with a clustering factor $\varepsilon = 0.0005$. a) Magnitude vs. time for ETAS California earthquakes from 1960 - present (9/2023). b) Omori-Utsu decay for 128 ETAS mainshocks having magnitudes $M \geq 5.5$, , $p$-value over the interval shown by the red dashed line is $p = 0.86$. c) Shows the optimized state variable as a function of time, an enlarged version of Figure 1d. d) Shows the Positive Predictive Value, PPV or Precision. Magenta curve is the PPV for the state variable shown in c), where the vertical axis is the threshold $T_H$. The cyan lines represent the PPV for 50 random time series. Mean of the time series is the solid black line, and $1\sigma$ confidence is shown as the dashed lines. e). Magenta curve is the corresponding self information $I$, equation (15), on the horizontal axis as a function of the threshold value $T_H$ on the vertical axis. Again, the cyan curves are the self-information for the ensemble of 50 random time series, with mean (solid black line) and $1\sigma$ confidence as the dashed lines.

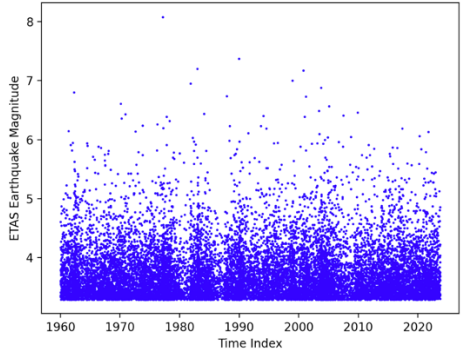

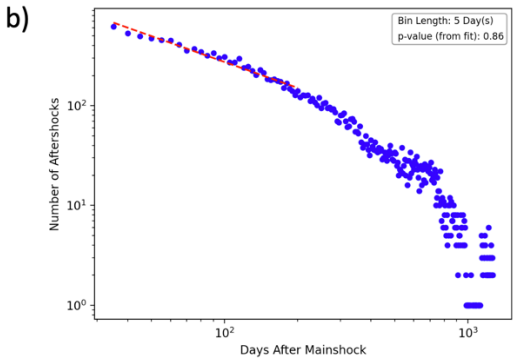

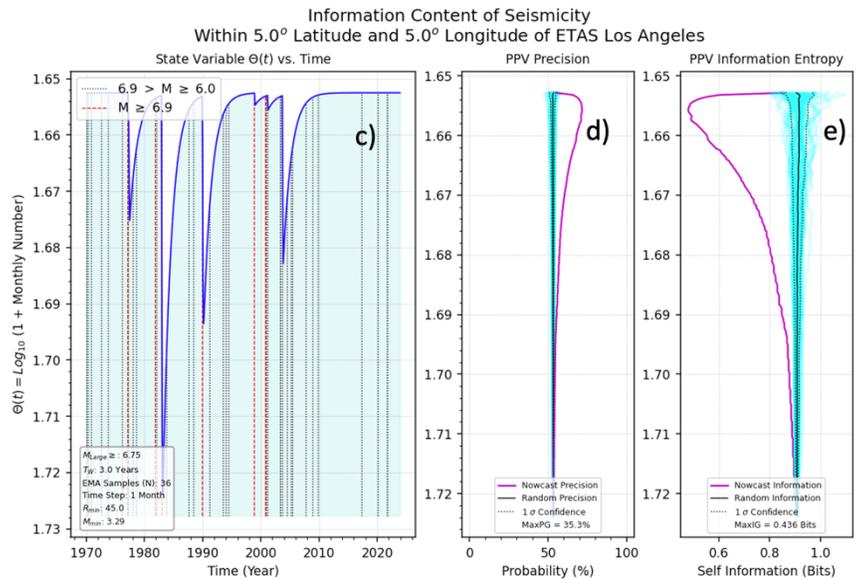

Small Non-Poisson ETAS Clustering with b = 0.95